\documentclass[aps,pra,reprint,showpacs]{revtex4-1}

\usepackage{amsmath,physics,graphicx,tabularx}

\begin{document}

\title{Cavity-enhanced photoionization of an ultracold rubidium beam for application in focused ion beams}

\author{G. ten Haaf}
\author{S.H.W. Wouters}
\author{P.H.A. Mutsaers}
\author{E.J.D. Vredenbregt}
\email{e.j.d.vredenbregt@tue.nl}
\affiliation{Department of Applied Physics, Eindhoven University of Technology, P.O. Box 513, 5600 MB Eindhoven, the Netherlands}

\date{\today}

\begin{abstract}
A two-step photoionization strategy of an ultracold rubidium beam for application in a focused ion beam instrument is
analyzed and implemented. In this strategy the atomic beam is partly selected with an aperture after which the
transmitted atoms are ionized in the overlap of a tightly cylindrically focused excitation laser beam and an ionization
laser beam whose power is enhanced in a build-up cavity. The advantage of this strategy, as compared to without the use
of a build-up cavity, is that higher ionization degrees can be reached at higher currents. Optical Bloch equations
including the photoionization process are used to calculate what ionization degree and ionization position distribution
can be reached. Furthermore, the ionization strategy is tested on an ultracold beam of $^{85}$Rb atoms. The beam
current is measured as a function of the excitation and ionization laser beam intensity and the selection aperture
size. Although details are different, the global trends of the measurements agree well with the calculation. With a
selection aperture diameter of 52 $\mu$m, a current of $\left(170\pm4\right)$ pA is measured, which according to
calculations is 63\% of the current equivalent of the transmitted atomic flux. Taking into account the ionization
degree the ion beam peak reduced brightness is estimated at $1\times10^7$ A/(m$^2\,$sr$\,$eV).
\end{abstract}


\maketitle

\section{Introduction \label{sec:Introduction_ionization}}

Photoionized ultracold atomic beams are promising to be used as source for focused ion beam (FIB) instruments for
nanofabrication purposes \cite{McClelland2016}. Atomic equivalent reduced brightnesses of the order of $10^7$
A/(m$^2\,$sr$\,$eV) have been reported \cite{Knuffman2013,tenHaaf2017_atombeam}, which are roughly an order of magnitude higher
than the industry standard liquid metal ion source (LMIS). Furthermore, energy spreads of less than 1 eV full width at
half maximum (FWHM) are expected \cite{Knuffman2013,tenHaaf2017_atombeam} as compared to 4.5 eV for the LMIS. Together with the
aberrations of the focusing column these two beam parameters determine the resolution of a FIB instrument for a given
current. With the aforementioned beam parameters and realistic focusing optics, a FIB probe size of 1 nm is expected
\cite{tenHaaf2014}.

Recently, two prototype ultracold cesium FIBs have been realized. Viteau et al. constructed an ion microscope based on
the field ionization of Rydberg-level promoted cesium atoms \cite{Viteau2016}. From the obtained resolution combined
with simulations an ion beam brightness of $2.8\times10^5$ A/(m$^2$ sr eV) was estimated. Furthermore, Steele et al.
realized a FIB system based on two-step but direct photoionization of cesium atoms \cite{Steele2017}. A direct
measurement of the ion beam's reduced brightness resulted in $2.4\times10^7$ A/(m$^2$ sr eV).

In this article the ionization strategy that is applied in the atomic beam laser cooled ion source (ABLIS)
\cite{Wouters2014} is introduced. In the ABLIS setup a thermal beam of Rb atoms effuses from a collimated
Knudsen source \cite{Wouters2016} and is laser cooled and compressed in the transverse direction. Subsequently the
atoms are ionized in a two-step photoionization process. This process distinguishes itself from other cold ion beam
sources by the use of an aperture to select the atoms to be ionized and a build-up cavity for the ionization laser,
which can bring advantages such as a higher ionization degree and faster ionization. The ions are created inside an electric
field, ranging from 10 kV/m to 1 MV/m, to suppress disorder-induced heating \cite{tenHaaf2014}. Therefore the energy
spread of the ions will be proportional to the range of positions at which they are produced, so the faster the atoms
are ionized the smaller the energy spread will be. After ionization the ions can be introduced in an electrostatic
focusing column to focus them to a small spot.

Analytical and numerical calculations of the ionization scheme and experimental results to confirm these
calculations are presented. The basis of the calculations are the so-called optical Bloch equations. They are extended
to include the ionization process and solved analytically (under idealized circumstances) and numerically to provide
expectations of what ionized fraction can be reached and in what longitudinal distance this happens as this distance
determines the energy spread of the ion beam. Measurements are performed of the beam current as a function of the
ionization laser beam intensity, excitation laser beam intensity and the selection aperture size in order to verify
some of the calculations.

Section \ref{sec:strategy_ionization} introduces the ionization strategy and discusses its advantages and disadvantages
in comparison with other ionization schemes. In section \ref{sec:equations_ionization} the optical Bloch equations
including ionization are introduced. Section \ref{sec:analytical_ionization} treats the analytical solution to these
equations under idealized circumstances, whereas section \ref{sec:numerical_ionization} treats the numerical solution
under realistic experimental circumstances. The experimental setup and results are discussed in section
\ref{sec:experiment_ionization}. Finally, section \ref{sec:conclusion_ionization} provides the conclusion of this work.

\section{Ionization strategy \label{sec:strategy_ionization}}

A commonly used scheme for the ionization of ultracold alkali atoms to form charged particle beams is two-step
photoionization \cite{McClelland2016}. An excitation laser beam excites the atoms to an intermediate state, from which
an ionization laser beam can bring the electron above the ionization threshold. Two lasers are used not only because
direct photoionization from the ground state would require laser frequencies in the UV-region but also because this
two-step photoionization scheme offers the possibility to define the ionization region by overlapping the foci of the
laser beams. Defining the ionization region serves two purposes. Firstly, the longitudinal size of the region
determines the energy spread, because the ionization usually takes place inside an electric field in order to preserve
the brightness. Secondly, the transverse size of the region influences the transverse current distribution. This
ionization strategy was used in the UCIS \cite{Debernardi2012} and MOTIS \cite{Knuffman2011} setups in which an
ionization volume was created inside a magneto-optical trap (MOT). Although in this way ion beams could be created with
extremely small energy spreads \cite{Reijnders2009}, the current that could be extracted was limited by the low
diffusion rate into the ionization region. To overcome this limitation, new generation cold ion sources are based on
photoionization of cold atomic beams \cite{Knuffman2013,tenHaaf2017_atombeam}. For example, a current of 20 pA was produced from
an ultracold cesium beam by overlapping laser beams with rms sizes of 2.25 $\mu$m and 1.75 $\mu$m \cite{Knuffman2013}.
By defocusing the laser beam this current could be increased, but at the cost of a lower transverse brightness and
higher energy spread due to the lower ionization laser intensity.

An other strategy to ionize ultracold atoms is to not photoionize the atoms by bringing the electron above the
ionization threshold, but to promote them to a high lying Rydberg state and subsequently let them travel into
a high electric field region in which they are field ionized \cite{Kime2013}. The advantage of this technique is the
fact that the cross section of Rydberg excitation is larger than direct photoionization, which relieves the necessity
of a high laser beam intensity. However, a disadvantage is that a more complex laser system is needed to lock the laser
to the Rydberg transition. The energy spread of the beam is in this case determined by the electric field gradient,
which determines the ionization probability distribution. In the first realization an energy spread of 2 eV was
obtained \cite{Viteau2016}.

\begin{figure}
	\includegraphics{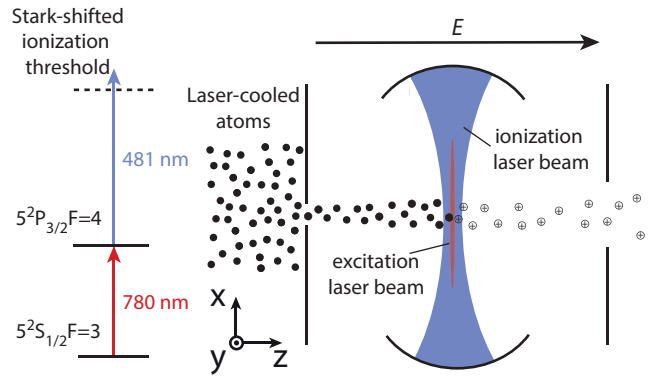}
	\caption{The strategy to ionize a beam ultracold $^{85}$Rb atoms. On the left a schematic energy level
	scheme is shown with the excitation and ionization transition indicated.  On the right a schematic side view of the ionization
	of atomic beam is shown. The ultracold rubidium atoms are selected with an aperture to set the ion beam current,
	excited by a tightly cylindrically focused excitation laser beam that travels in the $y$-direction and ionized by the
	overlapping ionization laser beam that travels in the $x$-direction, whose power is enhanced in a build-up cavity and
	which is focused less tightly than the excitation beam.}
	\label{photoionization_setup}
\end{figure}

Here, an alternative strategy to ionize an ultracold atomic beam is evaluated, which is schematically depicted in figure
\ref{photoionization_setup}. First of all, a selection aperture is used to select the part of the atomic beam that
is intended to be ionized. The atoms are then ionized in the crossover of an excitation laser beam and an ionization
laser beam. Because the cross section $\sigma_\text{i}$ for direct photoionization from the $5^2\text{P}_{3/2}$ state is
low a high laser intensity is needed to ionize this large fraction. As will be shown in subsection
\ref{subsec:simulation_ionization} an ionization laser intensity of more than $10^9$ W/m$^2$ is needed in order to
ionize more than 50\% of the incoming atoms. One option to realize this intensity is to focus the ionization laser beam
very tightly, as was done by Knuffman et al. to ionize their cesium beam \cite{Knuffman2013}. However, this tight waist
also limits the transverse size of the ionization region and therefore the maximum current that can be produced. An
other option is to increase the laser power by ionizing the atoms inside the ionization laser cavity
\cite{Gallagher1974} or an external build-up cavity to which the ionization light is coupled, the latter which is done
here. In this way the ionization laser beam can be focused less tightly to get the same laser beam intensity, which
enables a larger current at the optimal achievable brightness and energy spread. In order to limit the longitudinal
dimension of the ionization region the excitation beam is focused cylindrically and as tight as practically possible.

As mentioned, an advantage of the alternative strategy is the ability to create larger currents without sacrificing beam
brightness. A more practical advantage is that the current can be changed by simply using a different selection
aperture as is also done in LMIS based FIBs. Furthermore, when this selection aperture is placed close enough to the
ionization laser beam, it will define the transverse size of the ionization region. In that case ion beams with sharp
edges can be made, as compared to distributions with Gaussian-like tails. However, note that in the experiments discussed in
section \ref{sec:exp_setup_ionization} this is not yet the case as the aperture is placed too far from the ionization
position so that the transverse distribution is dominated by the temperature of the atoms. The main disadvantage of the
ionization strategy is the need for a build-up cavity and therefore a more complex design.

\section{Optical Bloch equations \label{sec:equations_ionization}}

In this section optical Bloch equations (OBE) are introduced that include the photoionization process in a standard way
\cite{Ackerhalt1977,Schek1985,Courtade2004,Deslauriers2006}. These equations are used in section \ref{sec:numerical_ionization} to
predict the ionization degree and the range over which ionization takes place in the ABLIS setup. The OBE are derived
from the so called master equation given by \cite{Rivas2012},
\begin{equation}
\frac{\partial\rho}{\partial
t}=\frac{i}{\hbar}\left[\rho,H\right]+\mathcal{L}\left(\rho\right),\label{master}
\end{equation}
in which $\hbar$ is the reduced Planck constant, $H$ is the Hamiltonian of the system under consideration,
$\mathcal{L\left(\rho\right)}$ is the Lindblad superoperator and $\rho$ is the density operator. The atom is
approximated by a three level system. The diagonal matrix elements $\rho_\text{gg}$, $\rho_\text{ee}$ and
$\rho_\text{ii}$ of the density operator are the populations of the ground, excited and ionized state respectively,
whereas the off-diagonal elements give the coherences between these states. By approximating the atom as a three level
system, the magnetic sub-level structures of the ground and excited state as well as states with a different total
angular momentum are not taken into account. The validity of this approximation is further discussed in subsection
\ref{subsec:discussion_ionization}.

The Hamiltonian accounts for the dynamics of the atom driven by the resonant excitation radiation field. In a rotating
frame and under the rotating wave approximation and the electric dipole approximation, it can be written as
\cite{Metcalf1999},
\begin{equation}
H=\hbar\begin{pmatrix}
0 & \Omega/2 & 0 \\
\Omega/2 & 0 & 0 \\
0 & 0 & 0
\end{pmatrix},
\end{equation}
in which $\Omega$ is the Rabi frequency which is given by,
\begin{equation}
\Omega=-\frac{\bra{e}\mathbf{E}\cdot\mathbf{d}\ket{g}}{\hbar}=\sqrt{\frac{s}{2}}\gamma,
\label{Rabi_equation}
\end{equation}
in which $\mathbf{E}$ is the excitation electric field amplitude, $\mathbf{d}$ is the electric dipole operator,
$s=\frac{I_\text{e}}{I_\text{s}}$ is the saturation parameter in which $I_\text{e}$ is the intensity of the excitation
radiation field and $I_\text{s}$ is the saturation intensity of the excitation transition. $\gamma$ is the linewidth of
the excitation transition, which in the case of rubidium from the $\mathrm{5^2S_{1/2}}$ state to the
$\mathrm{5^2P_{3/2}}$ state is $2\pi\times6.06$  MHz \cite{Steck2013}.

The Lindblad superoperator takes into account all interactions of the system with the environment and is given by
\cite{Rivas2012},
\begin{equation}
\mathcal{L}\left(\rho\right)=\frac{1}{2}\sum\limits_\text{j}\left(\left[V_\text{j}\rho,V_\text{j}^{\dagger}\right]+\left[V_\text{j},\rho
V_\text{j}^{\dagger}\right]\right),
\end{equation}
in which the summation is performed over all environment-induced transition operators $V_\text{j}$. In the problem
described here there are two relevant transition operators, namely the operators describing spontaneous emission
$V_\text{eg}$ and the ionizing transition $V_\text{ei}$, which are given by
\begin{equation}
\begin{split}
V_\text{eg}&=\sqrt{\gamma}\ket{g}\bra{e},\\
V_\text{ei}&=\sqrt{\gamma_\text{i}}\ket{i}\bra{e},\\
\end{split}
\end{equation}
in which $\gamma_\text{i}$ is the ionization rate from the excited state given by,
\begin{equation}
\gamma_\text{i}=\frac{\sigma_\text{i}I_\text{i}}{\hbar\omega_\text{i}},\label{ionrate}
\end{equation}
where $\omega_\text{i}$ and $I_\text{i}$ are the angular frequency and intensity of the ionization light and
$\sigma_\text{i}$ is the ionization cross section, which in the case of rubidium in the $\mathrm{5^2P_{3/2}}$ state is
$1.48\times10^{-21}$ m$^2$ \cite{Gabbani1997}. Note that dephasing due to the linewidth of the light field is not
included in the model. This is valid as long as this linewidth is much smaller than $\gamma$.

Combining equations \ref{master}-\ref{ionrate} results in a set of coupled linear differential equations for
$\rho_\text{gg}$, $\rho_\text{ee}$, $u=1/2\left(\rho_\text{ge}+\rho_\text{eg}\right)$ and
$v=i/2\left(\rho_\text{eg}-\rho_\text{ge}\right)$. Equations for the coherences involving the ionized state can also be
derived, but since these only influence each other and not the populations they are not of interest here. With the
introduction of $\mathbf{y}=\left(\rho_\text{gg},\rho_\text{ee},u,v\right)^\mathrm{T}$ the set of differential equations
can be written as,
\begin{equation}
\frac{\partial\mathbf{y}}{\partial t}=M\mathbf{y},\label{OBEmatrixform}
\end{equation}
in which $M$ is a matrix given by 
\begin{equation}
M=\begin{pmatrix}
0 & \gamma & 0 & -\Omega \\
0 & -\left(\gamma+\gamma_\text{i}\right) & 0 & \Omega \\
0 & 0 & -\frac{\gamma+\gamma_\text{i}}{2} & 0 \\
\frac{\Omega}{2} & -\frac{\Omega}{2} & 0 &
-\frac{\gamma+\gamma_\text{i}}{2}\\
\end{pmatrix}.\label{OBEmatrix}
\end{equation}
When equations \ref{OBEmatrixform}-\ref{OBEmatrix} are solved, $\rho_\text{ii}$ can be calculated with
\begin{equation}
\rho_\text{ii}=1-\rho_\text{gg}-\rho_\text{ee}.
\end{equation}

\section{Analytical solutions Optical Bloch equations\label{sec:analytical_ionization}}

There have been several reports on the analytical solutions of the optical Bloch equations or their magnetic
counterpart, describing nuclear magnetization dynamics under the influence of an external driving field
\cite{Bloch1946}. Although the latter describe a completely different physical system, the equations describing their
dynamics have the same structure. Ref. \cite{Torrey1949} provides an implicit solution to the original magnetic Bloch
equations which was later rederived explicitly \cite{Noh2010} to give solutions to the optical Bloch equations. A
solution was also found to the generalized magnetic Bloch equations describing the nuclear magnetization dynamics of a
decaying system \cite{Pottinger1985} which is a similar set of equations as equation \ref{OBEmatrixform}.

\begin{figure}
	\includegraphics{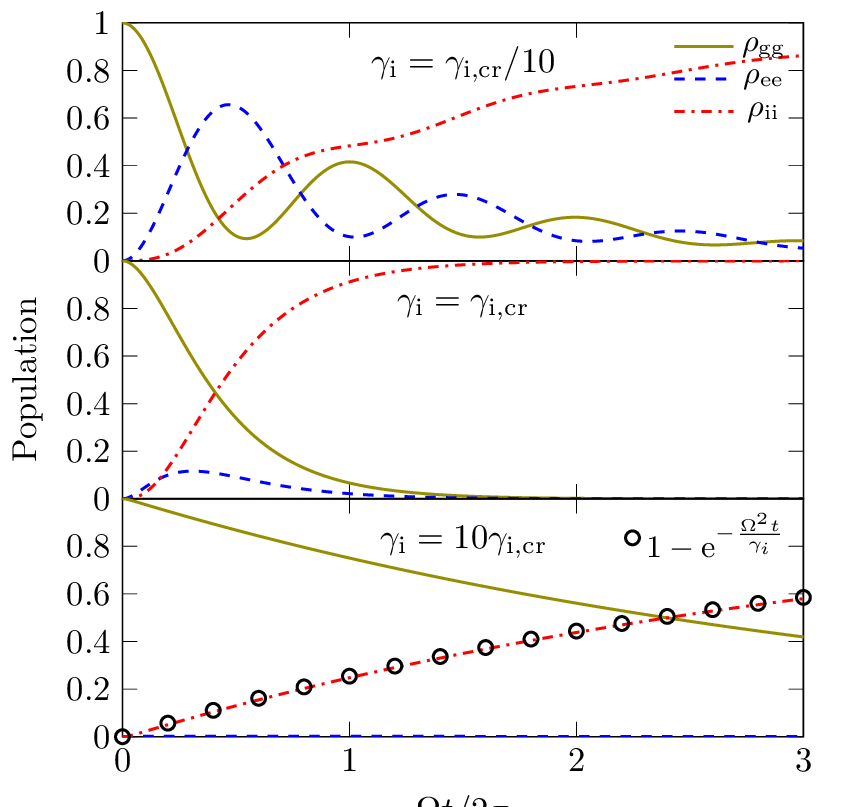}
	\caption{Analytical solutions to the OBE including ionization (equation \ref{OBEmatrixform}) in the case that
	$\gamma=\frac{\Omega}{10}$ and $\gamma_\text{i}=\frac{\gamma_\text{i,cr}}{10},\text{ }\gamma_\text{i,cr}\text{ and }
	10\gamma_\text{i,cr}$, where $\gamma_\text{i,cr}$ is the critical value for the ionization rate given by equation
	\ref{condition}. The solutions are given by equations \ref{case1} and \ref{case2}. For the case that
	$\gamma_i=10\gamma_\text{i,cr}$, a plot is also shown of the predicted ionized population by looking at the ionization
	as a continuous measurement of the excited state population which suppresses the excitation from happening by the
	quantum Zeno effect.}
	\label{analytical_example}
\end{figure}

An analytical solution to equation \ref{OBEmatrixform} is derived in the Appendix. The form of the solution depends on
the value of $\gamma_i$. Figure \ref{analytical_example} shows the solution for three cases:
$\gamma_\text{i}<\gamma_\text{i,cr}$, $\gamma_\text{i}=\gamma_\text{i,cr}$ and $\gamma_\text{i}>\gamma_\text{i,cr}$,
where the critical value $\gamma_\text{i,cr}$ is given by,
\begin{equation}
\gamma_\text{i,cr}=2\sqrt{\Omega^2+3\left(\frac{\gamma\Omega^2}{4}\right)^\frac{2}{3}}-\gamma.\label{condition}
\end{equation}
For each of these cases the populations of the ground state, excited state and ionized state are shown. In the first
case, Rabi oscillations between the ground and excited state can be seen, which are also characteristic for the
solutions to the original OBE. However, different in this case is that the Rabi oscillations are not only damped by
spontaneous emission, but also due to ionization which damps the oscillation to a zero equilibrium value of the excited
state population. In this regime the total ionization rate, which is the product of $\rho_\text{ee}$ and
$\gamma_\text{i}$, becomes higher for larger $\gamma_\text{i}$. In the second case the atom is almost completely in the
ionized state within the first Rabi period. The ionization is quenched within this time due to the fact that the
ionized state population is simply reaching 1. This is the situation in which the atom is ionized the fastest, because
when $\gamma_\text{i}$ is increased more, as shown in the third case, the total ionization rate becomes smaller. This
is caused by the fact that the Rabi oscillation is damped so fast that there is no reasonable excited state population
being developed and therefore also the ionized state is being populated slower. Therefore the larger the value of
$\gamma_\text{i}$ in this case, the longer it will take to reach the equilibrium ionized population $\rho_\text{ii}=1$.

The paradoxical behaviour that when $\gamma_\text{i}>\gamma_\text{i,cr}$, ionization happens slower for increasing
$\gamma_\text{i}$ can be explained as dissipative quantum Zeno dynamics \cite{Barontini2013}. The quantum Zeno effect
is the suppression of a transition between two quantum states due to the continuous measurement of one of the states
\cite{Itano1990}. Each time the occupation of one of the states is measured, the wave function collapses and if the
time between measurements is short enough the collapse is onto the original state. For the system described here,
the ionization process can be viewed as a continuous measurement of the excited state population. For continuous
measurements, at a rate $\gamma_i\gg\Omega$ this leads to an effective decay from the original state with a rate
$\frac{\Omega^2}{\gamma_i}$ \cite{Schulman1998,Streed2006}. That means that in this case the ionized population will
grow as $\rho_{ii}\left(t\right)\approx 1-\mathrm{e}^{-\frac{\Omega^2t}{\gamma_i}}$, which is indeed the case as can be seen in
the bottom panel of figure \ref{analytical_example}.


For a given $\Omega$ and $\gamma$ ionization happens fastest when $\gamma_\text{i}=\gamma_\text{i,cr}$ as discussed
above. As can be seen in figure \ref{analytical_example}, approximately 90\% is then ionized within a single Rabi
period. However, for a given $\gamma_\text{i}$, ionization will happen fastest for $\Omega\rightarrow\infty$. Taking
this limit in the analytical solution results in
$\rho_\text{ii}\left(t\right)=1-\text{e}^{-\frac{\gamma_\text{i}t}{2}}$, in which the factor $\frac{1}{2}$ in the
exponent can be explained by the fact that on average only half of the atoms is in the excited state. With this
equation and equation \ref{ionrate} the time it takes to get a certain ionized population can be calculated for the
case the excitation field is switched on instantaneously. However, in the experiment discussed here this is not the
case as the atoms travel through an excitation laser beam with a finite waist at a finite velocity. In order to find
the solution in this case a numerical approach is taken in the next section.

\section{Numerical solution Optical Bloch Equations\label{sec:numerical_ionization}}

In the previous section an analytical solution is found to the optical Bloch equations including ionization in the case
that the radiation fields are constant and switched on instantaneously. However, the atom travels through a Gaussian
shaped excitation laser beam, meaning that this field will change in amplitude gradually. To solve the optical Bloch
equations in this case a numerical approach is taken. Particularly of interest in the solutions are the probability
that an atom of the incoming beam is ionized after passing through the crossover of the excitation beam and ionization
beam and if the atom is ionized also the position at which this happens since this will determine the energy the atom
will gain from the acceleration field.

\subsection{Simulation setup}

The optical Bloch equations solve the population dynamics as a function of time. However, here the interest is in the
dynamics as a function of the longitudinal position, since this will determine the electric potential at which an ion
is made and thus the energy it will gain from the acceleration field. Furthermore the radiation field is dependent on
the position of the particle rather than on time. Therefore the time dependence in equation \ref{OBEmatrixform} is
changed to a position dependence by the transformation $t=z/v_z$, in which $z$ is the longitudinal position of the atom
measured with respect to the center of the excitation laser beam and $v_z$ is the longitudinal velocity of the atom.

For simplicity it is assumed the atoms are only moving in the $z$-direction, which is realistic as the atoms are
laser-cooled in the transverse direction. The longitudinal velocity is assumed to be distributed according a Maxwell
Boltzmann distribution. Measurements have shown this is not completely true \cite{tenHaaf2017_atombeam}. However, a Maxwell
Boltzmann distribution provides correct asymptotic behaviour and can be tuned to give the correct average velocity,
which is the case for a temperature of 22 K. The velocity distribution is discretized and for each of the
discretization steps the OBE are solved and the final solutions are weighted averages over these steps according the
velocity distribution.

As explained in section \ref{sec:strategy_ionization}, the ionization laser beam has a larger waist than the excitation
beam. In order to simplify the simulation and make it more comprehensible, since it releases the ionization laser beam
position as an extra simulation parameter, the ionization laser beam intensity distribution is assumed to be
independent of $z$. For simplicity, it is also assumed to be independent of $y$. The validity of this approach is
discussed in subsection \ref{subsec:discussion_ionization}. As stated in section \ref{sec:strategy_ionization} the
ionization laser power is enhanced with a build-up cavity. In the simplest case of a linear cavity this means the light
field of the ionization laser constitutes a standing wave in which the ionization light intensity $I(x)$ is given by
\begin{equation}
I\left(x\right)=4I_\text{i}\sin^2\left(\frac{2\pi x}{\lambda}\right),
\end{equation}
in which $\lambda$ is the wavelength of the ionization light and $I_\text{i}$ represents the intensity of the running
wave inside the cavity. To take this effect into account one period of the standing wave pattern is discretized. For
each of the intensities a simulation is performed and weighted averaging takes place over this intensity distribution.
Note that this is only valid if the size of the atomic beam is much larger than $\lambda/2$, which is usually the case
in realistic situations.

For the excitation laser beam a Gaussian intensity profile with a $1/\text{e}^2$ diameter of $4\sigma_\text{exc}=12$
$\mathrm{\mu m}$ is assumed in the longitudinal direction. It is assumed to be uniform in the transverse direction which
is a valid approximation since it will be formed by cylindrically focusing the excitation beam.

\subsection{Simulation results \label{subsec:simulation_ionization}}

\begin{figure}
	\includegraphics{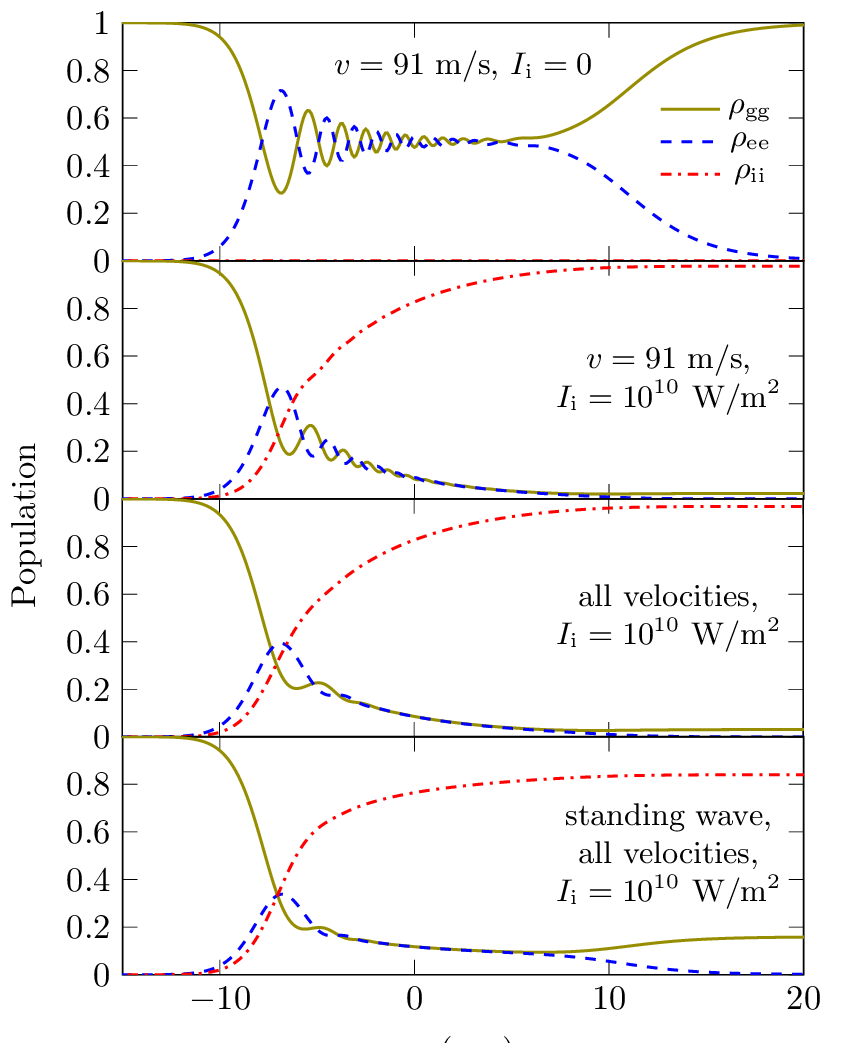}
	\caption{The ground state population ($\rho_\text{gg}$), excited state population ($\rho_\text{ee}$) and ionized state
	population ($\rho_\text{ii}$) as a function of position for four different situations: for a single atom traveling at
	91 m/s and without ionization, for a single atom traveling at 91 m/s and with ionization, for all atoms averaged over
	the longitudinal velocity distribution, and for all atoms averaged over the longitudinal velocity distribution and
	averaged over a standing wave of ionization light intensity.}
	\label{numerical_examples}
\end{figure}

Figure \ref{numerical_examples} shows the results of four simulations in which one by one several effects are taken into
account. The first plot shows simulation results in which an atom that travels through a Gaussian shaped excitation beam
with 91 m/s, but in which there is no ionization ($I_\text{i}=0$). The excited state population $\rho_\text{ee}$
undergoes a damped Rabi oscillation towards $\rho_\text{ee}=0.5$ in which the Rabi frequency $\Omega$ first increases
for $z<0$ and then decreases for $z>0$. As soon as $\Omega$ becomes smaller than the spontaneous damping rate $\gamma$
the excited state population undergoes an exponential decay. In the second plot $I_\text{i}=10^{10}$ W/m$^2$. Now an
ionized state population starts to develop as soon as there is an excited state population. There are still Rabi
oscillations in the excited state, but they are damped faster due to the additional damping by the ionization. Since
the atom gets ionized this damping is towards $\rho_\text{ee}=0$. In the third plot the same simulation is averaged over
the longitudinal velocity distribution. The trend of the simulations is very similar and the state populations in each
atom still undergo Rabi oscillations, even with the same frequency at each position. However, the difference is that
each atom travels with a different velocity and thus has acquired a different phase at each point. Therefore the
averaging washes out the Rabi oscillations in the final result. The final plot shows the effect of the standing wave of
ionization light. At the anti-nodes of the standing wave the populations will develop as in the third plot, but with a
slightly higher ionization rate due to the four times higher intensity as compared to a single running wave. However,
exactly at the nodes the populations will develop as in the first plot, i.e., no ionization will happen. Furthermore,
at points near the nodes the ionization will happen much slower. The average effect of this is that the final ionized
population is lower.

The two most interesting properties to look at in this research are the total ionized population after passing through
the laser beams and the ionization position distribution. These properties will influence the brightness and energy
spread of the ion beam. To find the ionization position distribution the ionized population as a function of position
that is found in the numerical calculation is differentiated with respect to the position. To further quantify the
results a full width at half maximum (FWHM) and a full width containing 90 \% of the particles (FW90) is calculated
from this distribution.

\begin{figure}
	\begin{tabular}{l}
			\includegraphics{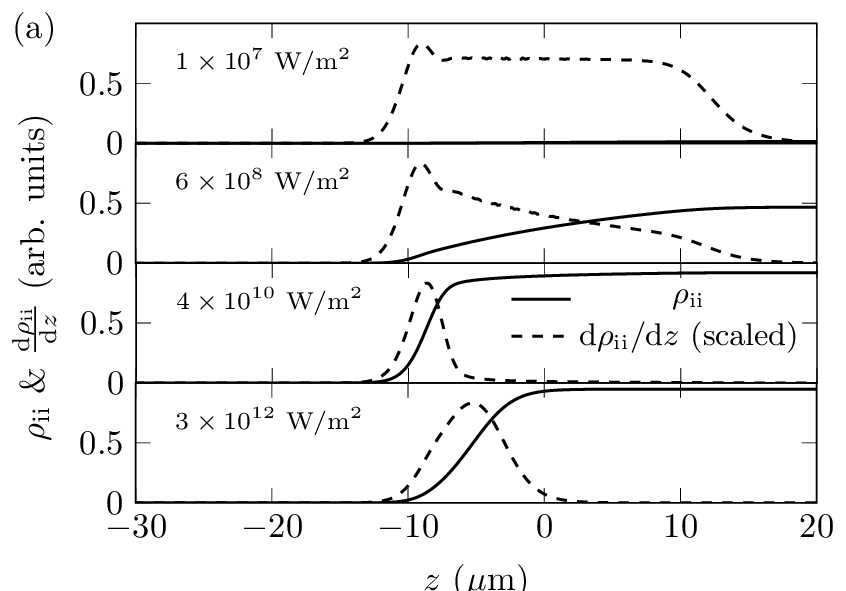}\\
			\includegraphics{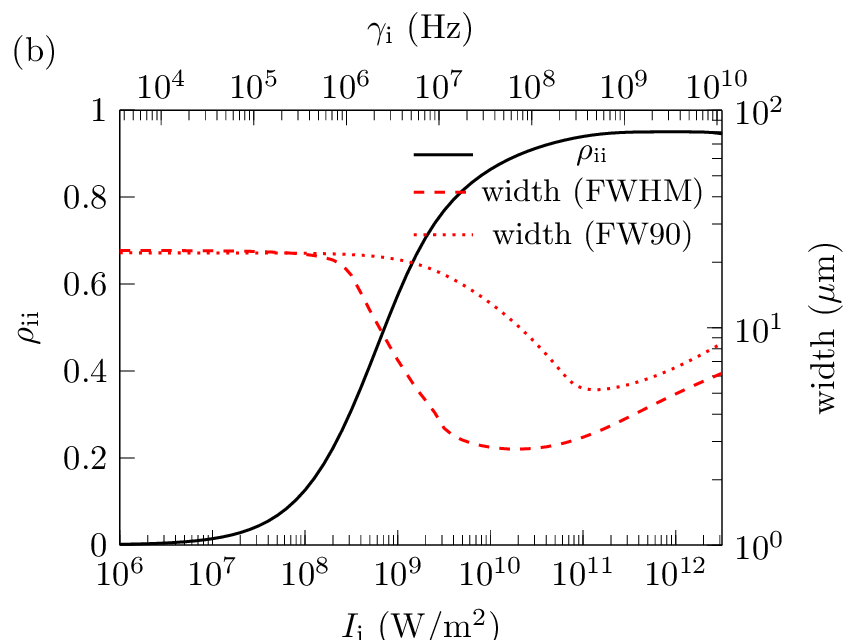}\\
	\end{tabular}
	\caption{Simulation results which show the dependence of the ionization process on the ionization laser intensity: (a)
	The ionized population and the ionization position distribution as a function of the longitudinal position at several
	different ionization laser intensities, indicated in the plots. (b) The ionization degree (corresponding to the left
	axis) and the FWHM and FW90 width of the ionization position distribution (corresponding to the right axis) as a function of
	the ionization laser intensity. Shown on the top axis is also the corresponding ionization rate. All simulations in this
	figure are done with $s=3000$ and $\sigma_{e}=3$ $\mu$m.}
	\label{ion_results}
\end{figure}

Figure \ref{ion_results} shows the dependence of the ionization process on the ionization laser intensity. Figure
\ref{ion_results}a shows some examples of the ionized population and the ionization position distribution as a function
of position. At very low ionization laser intensities no significant ionized population develops. In this regime the
ionization position distribution is completely determined by the shape of the excitation laser and the lifetime of the
excited state. At what position the ionization starts is determined by the intensity of the excitation light in
combination with the shape of the laser beam. When $\Omega\left(z\right)$ becomes larger than $\gamma$, a significant
excited state distribution starts to develop and thus ionization starts to occur. Then if $\Omega\left(z\right)$ again
drops below $\gamma$ there is an exponential decay in excited state population and thus also in the ionization position
distribution. As can be seen in figure \ref{ion_results}b this results in an ionization position distribution width of
roughly 22 $\mu$m (FW90).

A significant ionized population develops when $1/\gamma_\text{i}$ is of the same order as the average transit time of
an atom through the length in which there is a a significant excited state population. Dividing the average velocity by
a length of 22 $\mu$m gives an inverse transit time of 3 MHz. As can be seen in figure \ref{ion_results}b this agrees
with the value for $\gamma_\text{i}$ at which the ionization degree grows the fastest. From this value of
$\gamma_\text{i}$ the ionization position distribution also starts to become narrower because ionization is quenched
due to the fact that most of the atoms get ionized instead of the fact that there is no excitation from the ground
state anymore.

When the $I_\text{i}$ is so high that $\gamma_\text{i}$ becomes of the same order as the maximum Rabi frequency, which
is 1.5 GHz in this simulation, the ionization position distribution width reaches its smallest value. If $I_\text{i}$ is
increased more at this point the distribution becomes wider as a result of the fact that the Rabi oscillation becomes
overdamped as was also witnessed in the analytical solution. The smallest distribution width is, with a FWHM value of 3
$\mu$m, even smaller than the FWHM width of excitation laser beam, being 7 $\mu$m. Also note that $\rho_\text{ii}$ does
not get any higher than roughly 0.95 as a result of the standing wave distribution of ionization light. Because of the
standing wave, no matter how high the intensity becomes there are always atoms that experience a zero light intensity
at the nodes of the standing wave. This effect, in combination with the limited time an atom is excited, causes the
ionized population not to reach 1.

\begin{figure}
	\begin{tabular}{l}
			\includegraphics{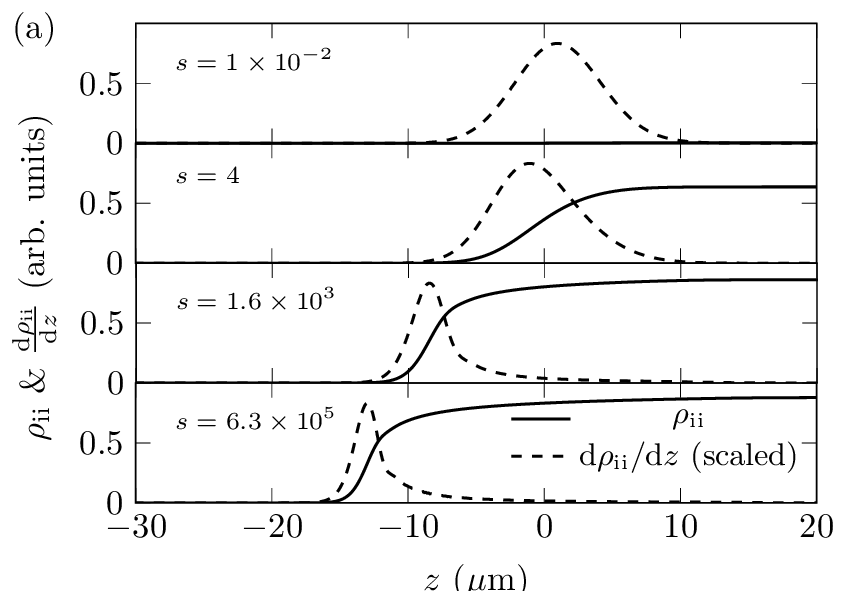}\\
			\includegraphics{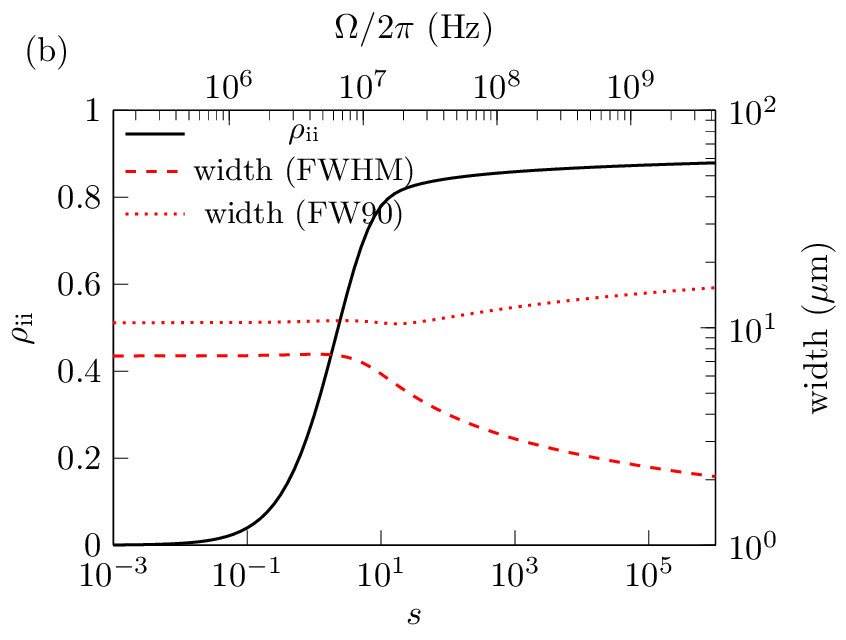}\\
	\end{tabular}
	\caption{Simulation results which show the dependence of the ionization process on the excitation laser intensity: (a)
	Ionized population and the ionization distribution as a function of the longitudinal position at several different 
	maximum excitation laser saturation parameters, indicated in the plots. (b) The ionization degree (corresponding to the
	left axis) and the FWHM and FW90 width of the ionization position distribution (corresponding to the right axis) as a
	function of the excitation laser saturation parameter. Shown on the top axis is also the corresponding Rabi frequency.
	All simulations in this figure are done with $I_\text{i}=10^{10}$ W/m$^2$ and $\sigma_{e}=3$ $\mu$m.}
	\label{exc_results}
\end{figure}

Figure \ref{exc_results} shows similar plots as in figure \ref{ion_results}, but in which $s$ is varied instead of
$I_\text{i}$. There is no significant ionization at low $s$, since there is no excited state population being developed.
This changes when $s$ becomes so large that $\Omega/2\pi$ becomes larger than $\gamma/2\pi$ and of the order of the
average inverse transit time of the atom through the laser beam. With a $1/\text{e}^2$ laser beam diameter of 12 $\mu$m
this average inverse transit time becomes 6 MHz which is indeed where a significant ionized populations starts to appear.
When $\Omega/2\pi$ is a few times the value of this inverse transit time, the total ionization degree does not grow
much anymore with an increase in intensity. What does happen is that the ions are created earlier since the local Rabi
frequency becomes larger than $\gamma$ earlier. As can be seen in figure \ref{exc_results}b this is accompanied by a
decrease in the FWHM of the ionization position distribution. Besides practical limitations, the limit to this decrease
in ionization position distribution width is that the ionization also has a finite rate. This is also the reason that
the FW90 value of the ionization position distribution is little affected. The reason why it increases is as follows. As
the Rabi frequency becomes larger the region in which there are excited atoms becomes larger and therefore, in the low
intensity part of the standing wave of ionization light, more atoms are ionized. These atoms are however ionized over a
larger range thus increasing the width of the ionization position distribution, which is mostly visible in the FW90 of
the distribution.

\subsection{Simulation discussion \label{subsec:discussion_ionization}}

Figure \ref{exc_results}b shows that increasing the Rabi frequency decreases the FWHM of the ionization position
distribution. However, this will only be true if the shape of the excitation field is perfectly Gaussian, i.e., if the
excitation beam also resembles a Gaussian far off-axis, since the ionization takes place further off-axis when the
excitation intensity is increased. In practice this is not the case. Even if the incoming excitation laser beam will be
perfectly Gaussian, the limited size of the focusing lens \cite{Saga1981} and aberrations of the lens will deform the
Gaussian intensity pattern far from its center. Therefore care should be taken in interpreting the results in figure
\ref{exc_results} where the ionization takes place far from the center of the excitation laser beam.

Another aspect to keep in mind is that in reality the system is not a real three level system. First of all, the ground
and excited state are in reality a multitude of magnetic sub-states, which each have a different transition strength an
thus a different saturation intensity. The distribution of the atoms over the magnetic sub-levels can be influenced by
optical pumping. For the results shown in figures \ref{ion_results} and \ref{exc_results} the average time an excited
state exists is roughly in the range 30-200 ns, depending on the values of $s$ and $I_i$. Since the lifetime of the
excited state is 26 ns this short interaction time means that only limited redistribution of the magnetic sub-levels
might occur, in contrast to ref. \cite{Porfido2015}. Furthermore, for the $F=3\rightarrow F=4$ excitation
transition discussed here, for which there are no dark magnetic sub-levels, the only effect will be that the effective
saturation intensity will be different and there will be some additional dephasing of the Rabi oscillations in the
total amount of population in an excited state.

Another approximations of the three level atom is that there are no other angular momentum states, whereas in reality
there exist multiple hyperfine levels. If the saturation parameter of the excitation laser beam becomes so high that the
Rabi frequency is of the same order as the detuning of the laser with respect to some other excited state, excitation
might also occur to this other state. However, if the ionization laser frequency is high enough so that ionization also
happens from this state, this will only add additional dephasing of Rabi oscillations. Furthermore, in the parameter
regime which is most suitable for practical application, most of the atoms are already ionized at the point where the
Rabi frequency is still smaller than the level splitting between between the intended excited level and the nearest
other one. For example, the ionization position distribution for $I_\text{i}=4\times10^{10}$ W/m$^2$ in figure
\ref{ion_results}a peaks at the point where the local Rabi frequency is about 25 MHz. This is still significantly
smaller than the hyperfine splitting of 120 MHz between the F=3 and F=4 state of the $\mathrm{5^2P_{3/2}}$ level in
$\mathrm{^{85}Rb}$.

In the numerical calculations the ionization intensity was assumed to be constant. However, in practice it will also
have a Gaussian beam shape. The effects of this ionization beam shape will be minor as long as the excitation and
ionization beam are overlapping well and the waist size of the ionization laser beam is significantly larger than the
region over which a significant excited state population exists.

\section{Experimental verification\label{sec:experiment_ionization}}

In order to verify some of the model calculations, measurements have been performed of the beam current as a function of
the excitation laser beam intensity, ionization laser beam intensity and the selection aperture size. These measurements
are compared with numerical calculations under the same conditions. In the next two sections the experimental setup and
the results are discussed.

\subsection{Experimental setup\label{sec:exp_setup_ionization}}

The beam of $^{85}$Rb atoms that is ionized in this experiment is the same as was described in previous work
\cite{tenHaaf2017_atombeam}. The measurements described here are performed at a magnetic field gradient of 1.1 T/m in
the 2D MOT and optimal laser cooling detuning. In the  measurements as a function of laser beam intensities the
temperature of the Knudsen source was 393 K, in the measurement as a function of aperture size the temperature was 433
K. From the previous work a total incoming flux equivalent to $\left(0.3\substack{+0.1\\-0.1}\right)$ nA and
$\left(0.6\substack{+0.3\\-0.2}\right)$ nA is expected at these conditions respectively. After the atomic beam
formation, the beam drifts 40 mm to the selection aperture. After the aperture the beam drifts for another 30 mm before
it reaches the laser beams and is ionized.

As explained in section \ref{sec:Introduction_ionization}, a build-up cavity is used to enhance the power in the
ionization laser beam. The build-up cavity consists of two concave mirrors placed inside the vacuum system. The $1/e^2$
diameter of the beam waist inside the cavity is 68 $\mu$m. The mirror substrate is coated with a dielectric interference
coating with a 99.7\% reflection coefficient at 480 nm, which gives a maximum theoretical laser beam power enhancement
in the cavity of 333$\times$. The ionization light is created by a Coherent Genesis MX 480 optically pumped semiconductor
laser, which produces a single mode beam of 481 nm light. The build-up cavity is locked to this laser with a piezo
actuator on one of the cavity mirrors by means of Pound-Drever-Hall frequency stabilization \cite{Drever1983}. The
laser beam is mode matched to the cavity with two lenses. The power enhancement in the cavity was determined by
measuring the power of the light transmitted through one of the mirrors. The maximum power enhancement measured is
$\left(2.0\pm0.2\right)\times10^2$. When the power in the ionization laser beam is varied this is done by changing the
power output of the laser.

The excitation light comes from a diode laser that is frequency stabilized with a frequency offset servo with respect to
the laser cooling laser which was stabilized to the $5\,^2\text{S}_{1/2}\text{F}=3$ to $5\,^2\text{P}_{3/2}\text{F}=4$
transition. The excitation light is focused cylindrically with an acylindrical lens with a focal length of 18 mm that
was positioned inside the vacuum system. The waist position of the excitation laser beam is overlapped with the atomic
beam by adjusting the position of a one meter focal length lens outside the vacuum in the direction of travel of the
laser beam, while looking at the atomic beam's laser-induced fluorescence imaged onto a CCD camera. The laser beam has
a $1/e^2$ diameter of 5.4 mm before focusing, this would lead to a diffraction limited $1/e^2$ waist diameter of 3.3
$\mu$m. However, waist measurements outside the vacuum to test the effect of shifting the beam's waist on the waist
size, in which the beam was perfectly centered on the lens, resulted in $1/e^2$ waist diameters of 7-12 $\mu$m,
depending on the lens position. This dependence and the fact that the beam might have traveled through the lens
off-center make it difficult to know the exact size of the waist in the measurements. Therefore the size of the Gaussian
excitation waist is varied in the simulation to find the best fit with the measured data.

The photoionization takes place in the middle between two electrodes which are separated by a 3 mm gap. The potential on
first electrode is set at 1 kV while the second is grounded, thus creating a field of $3\times10^5$ V/m. A commercial
Faraday cup is placed approximately 50 mm further. The collected current is measured with an electrometer. Each data
point shown is averaged over 50 measurements. At currents above 0.1 pA the average one standard deviation relative
fluctuation is 5\%. Below this current value the fluctuations are sometimes higher due to noise by other causes than
fluctuations in the beam.

\subsection{Experimental results}

\begin{figure}
	\includegraphics{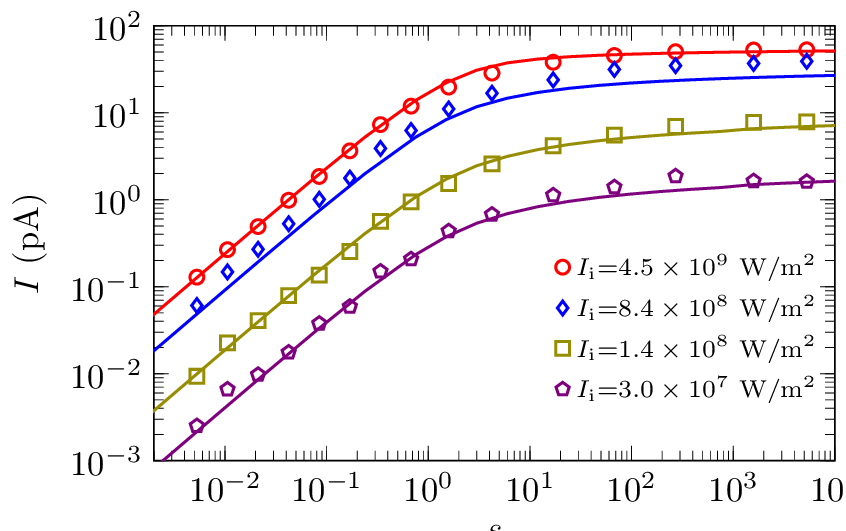}
	\caption{Experimental data (markers) of the beam current plotted against the excitation laser beam maximum saturation
	parameter for several different maximum ionization laser beam intensities. The atomic beam is selected with an
	aperture with a diameter of 52 $\mu$m. The lines represent simulation data under the same conditions as the measurement
	which were fitted to the measurement. The fit resulted in a maximum incoming flux equivalent to 81 pA and a saturation
	intensity of 52 W/m$^2$. The $\chi^2$ value of the fit is 11.}
	\label{exc_measurement}
\end{figure}

Figure \ref{exc_measurement} shows a measurement of the beam current as a function of the excitation beam maximum
saturation parameter $s$ for four different maximum ionization beam intensities $I_\text{i}$. In order to visualize the
dependence at low saturation parameters as well, the data is plotted on a double logarithmic scale. As can be seen the
current is proportional to $s$ when $s<1$. For $s\gg1$ the current saturates at a value which is higher for higher $I_\text{i}$. 

The saturation intensity used to calculate $s$ in figure \ref{exc_measurement} is found by comparing the experimental
results to the numerical calculations. The calculations are performed under the same conditions as the experiment is
done. In contrast to the simulations performed in section \ref{sec:numerical_ionization} also the limited spatial extent of the
ionization laser beam is included. This means that not only $\Omega$ is dependent on the longitudinal position of the
atom, but also $\gamma_\text{i}$. To include the dependence of $\gamma_\text{i}$ on the transverse position in the beam, an
additional weighted averaging step is included in the simulation over the transverse position, and corresponding
$I_\text{i}$, in which the weights are given by the transverse position distribution of the atoms. With the drift
distance of the beam after the selection aperture in this experiment (30 mm) this transverse position distribution is
not determined fully by the selection aperture, but by a combination of this aperture and the velocity distribution
of the atoms. In order to find out this distribution a Monte Carlo simulation is performed in which initial positions
and velocities after the laser cooling section are chosen according their known probability distributions. These
particles are then tracked to the selection aperture after which the selected particles are further tracked to the
ionization position. Also in contrast to the other simulations the results were averaged over the longitudinal velocity
distribution that was measured in \cite{tenHaaf2017_atombeam} instead of a Maxwell Boltzmann distribution. Furthermore, the
$1/e^2$ diameter of the excitation laser beam is varied to find the value that resulted in the lowest $\chi^2$ value of
the fit with the data.

The numerical calculations and experimental results are fitted to each other by finding the saturation intensity (to
fit the experimental data horizontally to the numerical data) and the current equivalent of the selected incoming
flux (to fit the numerical data vertically to the experimental data) that give the lowest $\chi^2$ value. The
uncertainties that are used to calculate this $\chi^2$ are the statistical fluctuations mentioned in section
\ref{sec:exp_setup_ionization}, systematic uncertainties are not taken into account. Since these fluctuations are smaller than the
markers of the experimental data in the figures in this section they are not displayed.

The lowest $\chi^2$ value is found for a $1/e^2$ diameter of the excitation laser beam of 16 $\mu$m, an incoming flux
equivalent to 81 pA and a saturation intensity of 52 W/m$^2$. The lines in figure \ref{exc_measurement} show the
fitted simulation data. The simulation overlaps well with the data points on the global scale. However, the $\chi^2$
value of the fit is 11, which indicates the model is not fully correct. Looking at the plots, this can be seen for
example by the fact that for $I_\text{i}=8.4\times10^8$ W/m$^2$ all data points are higher than the model, which
suggests that ionization is more efficient than predicted by the model at this ionization laser beam intensity. A
different ionization cross section can have been caused due to a dependence of the frequency of the ionization laser on
its power output. When the data for $I_\text{i}=8.4\times10^8$ W/m$^2$ is not taken into account a $\chi^2$ value of
5.6 is found, which indicates that a dependence of the ionization cross section on the ionization laser power can not
solely explain the difference between the calculation and the experimental data. A likely deviation of the model with
respect to the experiment is an imperfect Gaussian shape of the excitation laser beam waist. For low values of the
saturation parameter only the shape of the beam near the center is important, where the beam probably likely resembles
a Gaussian very well and the experimental data would thus agree with the model. However, at high saturation
parameters, the ionization takes place far from the center of the beam, as shown in figure \ref{exc_results}a. Here the
beam might not resemble a Gaussian shape very well, for example due to lens aberrations and finite apertures.

\begin{figure}
	\includegraphics{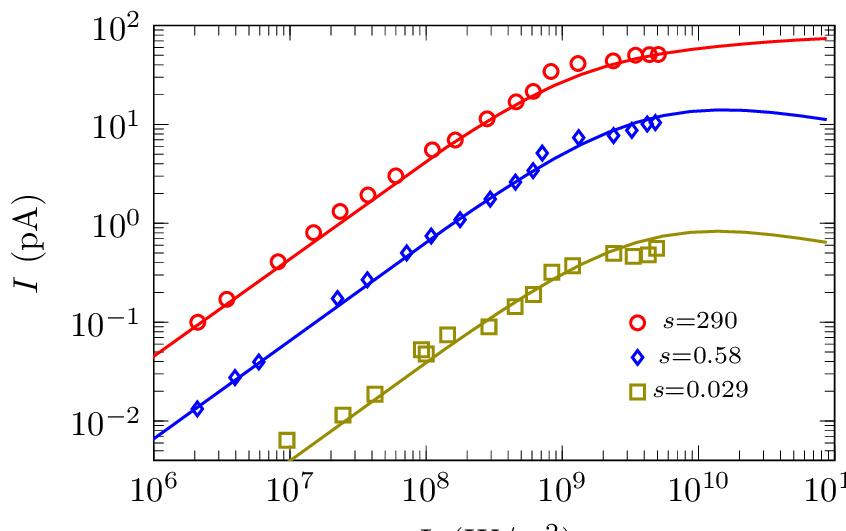}
	\caption{Experimental data (markers) of the beam current plotted against the ionization laser beam maximum intensity
	for several different maximum excitation laser beam saturation parameters. The atomic beam is selected with an
	aperture with a diameter of 52 $\mu$m. The lines represent simulation data under the same conditions as the
	measurement which were performed using the saturation intensity found from the data plotted in figure
	\ref{exc_measurement} and is also scaled with the same maximum incoming flux equivalent is found from that data.}
	\label{ion_measurement}
\end{figure}

Figure \ref{ion_measurement} shows experimental data of the beam current as function of $I_\text{i}$ for three
different values of $s$. The lines also show the corresponding calculation results. These results are calculated with
the same saturation intensity and scaled with the same maximum current as the results in figure \ref{exc_measurement}.
As can be seen the experimental data again overlaps well with the scaled calculation on a global scale, which confirms
the validity of the scaling values found from figure \ref{exc_measurement}. Note that, similarly as in the data shown
in figure \ref{exc_measurement}, there is a slightly higher current around $10^9$ W/m$^2$.

As said, a saturation intensity of 52 W/m$^2$ gives the best fit of the model to the experimental data. In the case
that the atoms are equally distributed over all magnetic sub-levels $m_F$ and are excited with linearly polarized
light, the saturation intensity would be 38.9 W/m$^2$ \cite{Steck2013}. The higher value found can be caused by a
magnetic sub-level distribution that has a higher probability for higher $|m_F|$, when excitation is done with linearly
polarized light.

\begin{figure*}
	\includegraphics{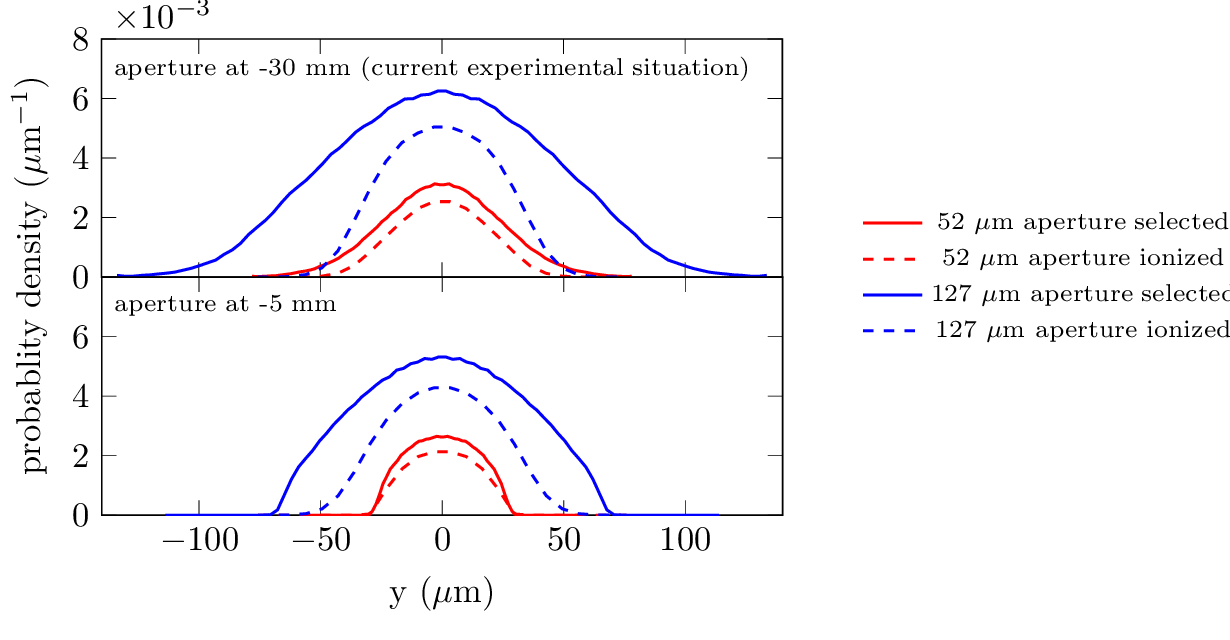}
	\caption{Transverse atom distribution just before ionization (solid curves) and transverse ion distribution just
	after ionization (dashed curves) as a function of $y$ (for orientation see figure \ref{photoionization_setup}) for two
	different aperture diameters (see legend) and in the situation that the aperture is positioned at $z=-30$ mm (top) and
	$z=-5$ mm (bottom). The distributions are normalized such that the integral results in the fraction of incoming atoms
	that is transmitted by the aperture or ionized.}
	\label{transverse_distributions}
\end{figure*}

The maximum measured current using the selection aperture with a diameter of 52 $\mu$m is 51 pA. Therefore the maximum
achieved ionization degree is 63\%. This is lower than the ionization degree of 81\% displayed in figure
\ref{ion_results} at a laser intensity of $5\times10^9$ W/m$^2$. The reason for this difference is the fact that the
transverse ionization position distribution in the experiment is wider than the 1/e$^2$ ionization laser beam waist of
68 $\mu$m. The transverse position distribution of the selected atomic beam at the position of the ionization, as is
found from the Monte Carlo simulation introduced earlier, is shown in the top panel of figure
\ref{transverse_distributions} for a selection aperture diameter of 52 $\mu$m and 127 $\mu$m. The figure also shows
$y$-position distribution of the ions just after the ionization (see figure \ref{photoionization_setup} for a
orientation reference). Note that the incoming flux was normalized to 1, and the distributions in figure
\ref{transverse_distributions} are scaled accordingly. Therefore the integral of the selected atom distribution results
in the probability that an atom in the incoming atomic beam is transmitted by the aperture and the integral of the ion
distribution results the probability that an atom in this incoming atomic beam is ionized. The ionization degree is
81\% in the center, while it is almost zero at 60 $\mu$m from the center. As can be seen the distributions are
significantly wider than the respective selection aperture diameters. This is caused by the finite temperature of the
atoms and the fact that they drift for 30 mm after the selection aperture in the current configuration. This
configuration can be improved on by placing the aperture closer to the ionization position. The bottom panel of figure
\ref{transverse_distributions} shows the results of a calculation of the same distributions when the aperture is placed
at $z$=-5 mm. As can be seen the distributions are narrower and have a sharper edge. If such a sharp-edged distribution
could be imaged onto the sample in a FIB instrument without artefacts, this will bring an advantage concerning
nanofabrication purposes. However, since this makes the design more challenging this configuration is not pursued in
this proof-of-concept experiment.

\begin{figure}
	\includegraphics{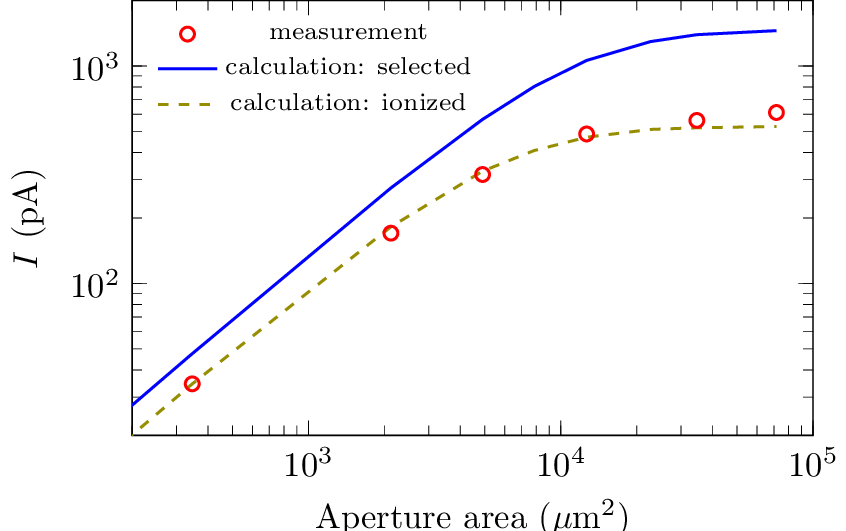}
	\caption{Experimental data (markers) of the beam current plotted against the area of the aperture used to select the
	atomic beam. The data is taken at a Knudsen source temperature of 433 K, with $I_\text{i}=6\times10^9$ W/m$^2$ and
	$s=6000$. The lines show simulation data under the same conditions as the measurement. The solid line shows the
	current equivalent of the atomic flux that is transmitted through the selection aperture. The dashed line shows the
	current that is actually created. A fit is performed to find the current equivalent of the (unselected) incoming flux
	that matches the calculation data best with the experimental data, which resulted in 1.4 nA.}
	\label{apertures_measurement}
\end{figure}

Figure \ref{apertures_measurement} shows the current as a function of the selection aperture area. The markers show the
experimental data, the dashed line shows scaled calculation data of the ionization degree and the solid line shows
scaled calculation data of the fraction of the incoming atomic flux that is transmitted by the selection aperture. As
this data is measured at a different Knudsen source temperature than the data in figures \ref{exc_measurement} and
\ref{ion_measurement} a new scaling is performed. The current equivalent of the incoming flux resulting from this
scaling is 1.4 nA, which is significantly larger than the value of $\left(0.6\substack{+0.3\\-0.2}\right)$ nA found
earlier by means of laser-induced fluorescence \cite{tenHaaf2017_atombeam}. With the new determined value of the current
equivalent of the incoming atomic flux, the ionization degree and the earlier measured temperature \cite{tenHaaf2017_atombeam}
of the incoming atoms the peak brightness of the produced ion beams is calculated to be $1\times10^7$
A/(m$^2\,$sr$\,$eV). By using a selection aperture with a diameter of 300 $\mu$m, the maximum produced current is
$\left(0.61\pm0.01\right)$ nA.

\section{Conclusion\label{sec:conclusion_ionization}}

In this work an alternative strategy to ionize an ultracold atomic beam is analyzed. In this strategy an aperture is
used to select the part of the ultracold beam that is to be ionized. This selection of atoms is ionized in two steps by
a tightly cylindrically focused excitation laser beam that is positioned within an ionization laser beam which
intensity is enhanced by a build-up cavity. With this strategy a high ionization degree over a larger transverse area
can be achieved as compared to without a build-up cavity, due to the high intracavity power. Therefore higher ion beam
currents can be reached at the maximum available brightness which is determined by the incoming atomic beam parameters.

The optical Bloch equations are extended to include the ionization process. These equations are solved analytically
with constant light fields. Although this analytical solution provides important insight the ionization dynamics,
especially in the case that $\gamma_\text{i}\gg\Omega$, a numerical solution is found with the experimentally achievable
laser fields taken into account. These provide expectations for the ionization degree and ionization position
distribution. The calculations show that it is possible to limit the longitudinal region of ionization by solely
focusing the excitation laser beam tightly. High ionization degrees can be reached when $\Omega/2\pi$ and the
$\gamma_\text{i}$ are both of the order of the inverse transit time of the atom through the excitation laser beam. An
ionization degree of 80\% is possible with an ionization intensity of $10^{10}$ W/m$^2$.

Experiments are performed in which the beam current is measured as a function of the excitation laser beam intensity
ionization laser beam intensity and the aperture size. Numerical calculations are done under similar conditions and
fitted with the experimental results in order to determine the current equivalent of the incoming flux of atoms and
the saturation intensity of the excitation transition. With the best fitting parameters the overall trend of the
measurement is reproduced. However, the $\chi^2$ value of the fit is 11, which indicates the model is not fully
correct. A likely difference between modeled and experimental situation is that the excitation laser beam waist did
not have a perfect Gaussian shape in the experiment. Also, the $\chi^2$ value is calculated while only taken into
account the statistical fluctuations in the current and not systematic uncertainties in for example the aperture size
and ionization cross section. With a selection aperture diameter of 52 $\mu$m, a maximum current of
$\left(170\pm4\right)$ pA is measured which according to simulation data is 63\% of the current equivalent of the
transmitted atomic flux. Using a larger aperture the maximum current produced is $\left(0.61\pm0.01\right)$ nA. Taking
the ionization degree into account the brightness of the produced ion beam is estimated at $1\times10^7$
A/(m$^2\,$sr$\,$eV).

\begin{acknowledgments}
This work is part of the research programme Ultracold FIB with project number 12199, which is (partly) financed
by the applied and engineering sciences division of the Netherlands Organisation for Scientific Research (NWO). The
research is also supported by Thermo Fisher Scientific, Coherent Inc. and Pulsar Physics.
\end{acknowledgments}

\appendix*
\section{Derivation analytical solution}

To find a solution to equation \ref{OBEmatrixform} a trial solution of the form
$\mathbf{y}=\mathbf{y_j}\mathrm{e}^{\lambda_\text{j} t}$ is inserted, which reduces the equation to the eigenvalue
problem
\begin{equation}
M\mathbf{y_j}=\lambda_\text{j}\mathbf{y_j}.\label{eigenvalueproblem}
\end{equation}
In order to find a non-trivial solution to this equation the following condition must hold:
\begin{equation}
\mathrm{det}\left(M-\lambda_\text{j} I\right)=0 \label{generalsolution},
\end{equation}
in which $I$ is the unity matrix. Substitution of equation \ref{OBEmatrix} and writing out the determinant results in
\begin{align}
	&\left(\lambda_j+\frac{\gamma+\gamma_\text{i}}{2}\right)\times\nonumber\\
	&\left(\lambda_j^3+\frac{3\left(\gamma+\gamma_\text{i}\right)}{2}\lambda_j^2+\left(\frac{\left(\gamma+\gamma_\text{i}\right)^2}{2}+\Omega^2\right)\lambda_j+\frac{\gamma_\text{i}\Omega^2}{2}\right)=0.\nonumber\\
	\label{eigenvalue_equation}
\end{align}
This equation shows that the first eigenvalue is,
\begin{equation}
\lambda_1=-\frac{\gamma+\gamma_\text{i}}{2},
\end{equation}
with its matching eigenvector,
\begin{equation}
\mathbf{y}_1=\left(0,0,1,0\right)^T.
\end{equation}
The other eigenvalues are the roots of the third order polynomial given by the second part of the left hand side of
equation \ref{eigenvalue_equation}. These roots are given by,
\begin{equation}
\begin{split}
\lambda_2&=-\alpha-2p,\\
\lambda_{3(4)}&=-\alpha+p+(-)iq,
\end{split}
\label{lambdas}
\end{equation}
in which,
\begin{equation}
\begin{split}
\alpha&=\frac{\gamma+\gamma_\text{i}}{2},\\
p&=\frac{1}{2}\left(\frac{Q}{S}+S\right),\\
q&=\frac{\sqrt{3}}{2}\left(\frac{Q}{S}+S\right),\\
S&=\left(R+\sqrt{R^2-Q^3}\right)^\frac{1}{3},\\
Q&=\frac{1}{12}\left(\left(\gamma+\gamma_\text{i}\right)^2-4\Omega^2\right),\\
R&=-\frac{\gamma\Omega^2}{4}.
\end{split}
\label{pqSQR}
\end{equation}
The values of these eigenvalues, especially whether they are purely real or complex is important since they determine
the time evolution of the density matrix components. A real eigenvalue means an exponential decay or growth, while
a complex eigenvalue indicates there is an oscillation in the density matrix components. Therefore the argument of the
square root in the equation for $S$ plays an important role. When $R^2>Q^3$, $p$, $q$ and $S$ are all purely real
quantities which means that $\lambda_3$ and $\lambda_4$ have complex values. However, if $R^2<Q^3$ it turns out that
$\lambda_3$ and $\lambda_4$ are purely real quantities, although this is not immediately clear from equations
\ref{lambdas} and \ref{pqSQR}. Therefore it is in this case better to work out the equations for $p$ and $q$, which
results in,
\begin{equation}
\begin{split}
p&=\sqrt{Q}\cos\left(\frac{1}{3}\arccos\left(\frac{R}{Q^\frac{3}{2}}\right)\right)=\sqrt{Q}\cos{\phi},\\
q&=i\sqrt{3Q}\sin\left(\frac{1}{3}\arccos\left(\frac{R}{Q^\frac{3}{2}}\right)\right)=i\sqrt{3Q}\sin{\phi}.\\
\end{split}
\label{pq2}
\end{equation}
Inserting these values in equations \ref{lambdas} shows that all eigenvalues become purely real.

The next step is to calculate the eigenvectors that match with the eigenvalues from equation \ref{lambdas}, which is
done by inserting the eigenvalues back into equation \ref{eigenvalueproblem} and solving for the eigenvectors. The
results are given by,
\begin{equation}
\mathbf{y}_\text{i}=\left(1+\frac{2}{\Omega^2}\left(\lambda_\text{i}+\alpha\right)\left(\lambda_\text{i}+2\alpha\right),1,0,-\frac{1}{\Omega}\left(\lambda_\text{i}+2\alpha\right)\right)^T,
\label{eigenvectors}
\end{equation}
which is valid for $i=2,3,4$. The complete solution to equation \ref{OBEmatrixform} is now given by
\begin{equation}
\mathbf{y}\left(t\right)=\sum\limits_{i=1}^4c_\text{i}\mathbf{y}_\text{i}\mathrm{e}^{\lambda_\text{i} t},\label{fullsolution}
\end{equation}
in which the coefficients $c_\text{i}$ are determined by the begin conditions. In the case the atom starts in the ground
state, i.e., $\mathbf{y}\left(0\right)=\left(1,0,0,0\right)^T$, the coefficients are given by 
\begin{equation}
\begin{split}
c_1&=0,\\
c_2&=\frac{\Omega^2}{2\left(9p^2+q^2\right)},\\
c_{3\left(4\right)}&=\frac{\Omega^2}{4q\left(-q+\left(-\right)3ip\right)}.\\
\label{coefficients}
\end{split}
\end{equation}
Combining equations \ref{lambdas}-\ref{fullsolution} gives you the complete solution to the problem. To make the
solution as clear as possible it is best to write down two solutions, one for $R^2>Q^3$ and one for $R^2<Q^3$. In the
first case, that can also be written as $\gamma_\text{i}<\gamma_\text{i,cr}$, where
$\gamma_\text{i,cr}=2\sqrt{\Omega^2+3\left(\frac{\gamma\Omega^2}{4}\right)^\frac{2}{3}}-\gamma$, the solution is given
by
\begin{align}
\mathbf{y}=&c_2\mathbf{y}_2\mathrm{e}^{-\left(\alpha+2p\right)t}+\nonumber\\
&2\mathrm{Re}\left(c_3\mathbf{y}_3\right)\cos\left(qt\right)\mathrm{e}^{-\left(\alpha-p\right)t}-\nonumber\\
&2\mathrm{Im}\left(c_3\mathbf{y}_3\right)\sin\left(qt\right)\mathrm{e}^{-\left(\alpha-p\right)t},\nonumber\\
\label{case1}
\end{align} 
in which $p$ and $q$ are given  by equation \ref{pqSQR} and for which is made use of the fact that $c_3=c_4^*$ and
$\mathbf{y}_3=\mathbf{y}_4^*$. In the other case all eigenvalues are real and the solution may be written as
\begin{align}
\mathbf{y}=&c_2\mathbf{y}_2\mathrm{e}^{-\left(\alpha+2\sqrt{Q}\cos\phi\right)t}+\nonumber\\
&c_3\mathbf{y}_3\mathrm{e}^{\left(\sqrt{3Q}\sin\phi\right)t}\mathrm{e}^{-\left(\alpha-\sqrt{Q}\cos{\phi}\right)t}+\nonumber\\
&c_4\mathbf{y}_4\mathrm{e}^{-\left(\sqrt{3Q}\sin\phi\right)t}\mathrm{e}^{-\left(\alpha-\sqrt{Q}\cos{\phi}\right)t},\nonumber\\
\label{case2}
\end{align}
in which $\phi$ is used which is defined in equation \ref{pq2}.

\bibliography{ionization}

\end{document}